\documentclass[prb,twocolumn,showpacs,preprintnumbers,amsmath,amssymb]{revtex4}
\usepackage{graphicx}
\usepackage{dcolumn}
\usepackage{bm}
\begin{document}

\preprint{APS/123-QED}

\title{
Theoretical evidences for enhanced superconducting transition temperature
of CaSi$_2$ in a high-pressure AlB$_2$ phase
}

\author{
A. Nakanishi, T. Ishikawa, H. Nagara, and K. Kusakabe
}

\affiliation{
Division of Frontier Materials Science, Graduate School of
Engineering Science, Osaka University, Toyonaka, Osaka 560-8531, Japan 
}

\date{\today}

\begin{abstract}
By means of first-principles calculations,
we studied stable lattice structures and 
estimated superconducting transition temperature
of CaSi$_2$ at high pressure.
Our simulation showed stability of
the AlB$_2$ structure 
in a pressure range above 17\,GPa.
In this structure,
doubly degenerated optical phonon modes,
in which the neighboring silicon atoms oscillate 
alternately in a silicon plane, show prominently
strong interaction with the conduction electrons.
In addition there exists a softened optical mode 
(out-of-plan motion of silicon atoms), whose strength of the
electron-phonon interaction is nearly the same as the above mode.
The density of states at the Fermi level in the AlB$_2$ structure
is higher than that in the trigonal structure.
These findings and the estimation of the
transition temperature strongly suggest that higher
$T_{\rm c}$ is expected in the AlB$_2$ structure
than the trigonal structures which are known so far.
\end{abstract}

\maketitle

\section{Introduction}

The calcium di-silicide, CaSi$_2$, has a rhombohedral crystal structure at 
the ambient pressure. 
In this layered structure, superconductivity does not appear
down to 0.03\,K.\cite{Phase-I} 
In a high pressure phase of the $\alpha$-ThSi$_2$ type structure, 
which is tetragonal,
CaSi$_2$ is known to become a superconductor with
the superconducting transition temperature $T_{\rm c}=1.58$\,K.\cite{Phase-II} 
At the pressure of $P\simeq$ 10\,GPa, CaSi$_2$ in the rhombohedral structure 
undergoes a structural phase transformation into another trigonal structure 
(Phase III) and superconductivity appears with $T_{\rm c}$ 
rising up to about 3\,K.\cite{sc,str}
In this structure, silicon atoms form 
corrugated honeycomb networks.
Between two Si honeycomb planes, 
Ca atoms are intercalated, forming another plane of a triangular lattice. 
Each Ca atom locates just above the center of one corrugated hexagon 
formed by six Si atoms in the adjacent Si plane. 
If the pressure increases up to 
$P\simeq$15\,GPa, another phase transformation takes place (Phase IV) and
the corrugated Si planes become nearly flat.
This phase is called Phase IV. The atomic structure of Phase IV is 
rather close to the AlB$_2$ structure, 
which has perfectly flat silicon planes. Since corrugation remains 
in Phase IV, it is
sometimes called the AlB$_2$-like structure.\cite{sc,str}
In this structure, $T_{\rm c}$ further rises up to around 14\,K, which is the 
highest record in CaSi$_2$.
Here, we note that before these experimental findings, the 
structural transition from the trigonal structure to the AlB$_2$ structure 
has been predicted theoretically by one of the authors 
of this study and his coworkers.\cite{KK}

The superconductivity in the AlB$_2$ structure attracted 
much attention after 
finding of a high-temperature superconductor. MgB$_2$,\cite{NNMZA}
As for CaSi$_2$, theoretical studies were done 
in low-pressure phases.\cite{Weaver,Fahy} 
Satta {\it et al.} considered possibility of the AlB$_2$ structure 
in a high pressure condition, but they could not find this structure 
with fixed cell parameters.\cite{Satta} 
The electron-phonon interaction and the superconducting transition 
temperature were rarely estimated theoretically.\cite{Wang} 
Thus, 
CaSi$_2$ has not been studied so often in the literature compared with MgB$_2$.
This another superconductor, however, has its own interest and importance. 
This is because CaSi$_2$ provides us with an ideal testing ground
on which we can compare several polymorphs showing superconductivity. 
Some of these polymorphs resemble each other,
but the superconducting transition temperature $T_{\rm c}$ changes its value 
when the structural phase transition takes place.
If we could find a key factor determining the change of $T_{\rm c}$
with the structural phase transition,
it would help us to understand 
this superconducting Zintl-phase compound. 

The purpose of the present study is to clarify the nature of 
CaSi$_2$ in high pressure phases above 10GPa. 
Using the first-principles calculations, 
we did the optimization of the atomic structures in CaSi$_2$ 
at high pressures.
We obtained an indication of the pressure-induced phase transition 
from the known trigonal structure to another high-pressure phase with 
the AlB$_2$ structure. 
Then we studied the electronic structures, the phonon dispersion relations,
and the superconducting transition temperature. 
Those results were compared with those of MgB$_2$.
We estimated the superconducting transition temperature by means of
the strong-coupling theory using the electron-phonon coupling
constants obtained by the first-principles calculations.
Our results tell us that CaSi$_2$ undergoes a structural phase
transformation to the phase of the AlB$_2$ structure 
and that the new phase is expected to be a superconducting phase
with much higher transition temperature 
than that of the trigonal structure phase. 

\section{Calculation Methods}

In this study, we consider the pressure range of $P=10\sim20$\,GPa. 
For the structural optimization, we started simulations from
lattices whose unit cell contain one Ca atom and two Si atoms
and without any symmetry requirements.
The space-group of the trigonal lattices of CaSi$_2$ is $P\bar{3}m1$
and that of AlB$_2$ structure is $P6/mmm$. 
The Wyckoff position of the calcium atom at the $1a$ site of $P\bar{3}m1$ is 
given by $(0,0,0)$, while those of two silicon atoms are $(1/3,2/3,z)$
and $(2/3,1/3,\bar{z})$ with the internal parameter $z$. 
When $z=0.5$, the structure becomes identical to the 
AlB$_2$ structure.

For the determination of the electronic structure,
we utilized the density-functional theory\cite{HK,KS} in the generalized
gradient approximation \cite{PW91} with the ultra-soft pseudopotentials for the atomic
potential.\cite{VB}
The wave functions and the electronic charge density 
are expanded in the plane-wave basis. 

For the pseudopotentials, 3s, 3p and 4s electrons of the calcium atom
are treated as the valence electrons,
and for silicon atom 3s and 3p electrons are used.
Calculations are performed by the use of software package, the Quantum ESPRESSO.\cite{PWSCF}

\begin{widetext}
\begin{figure}[htpd]
 \begin{center}
   \begin{tabular}{ccc}
    \resizebox{60mm}{!}{\includegraphics{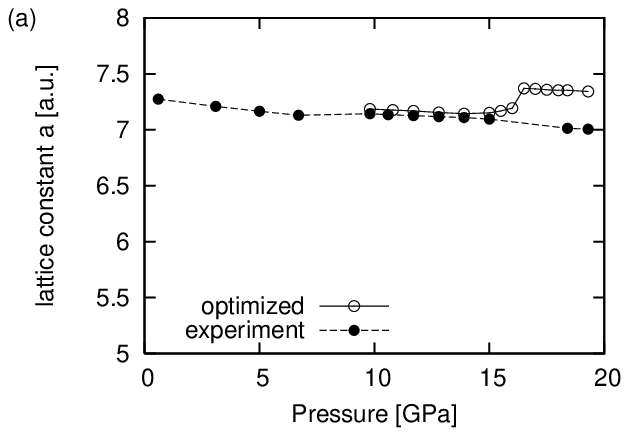}} &
    \resizebox{60mm}{!}{\includegraphics{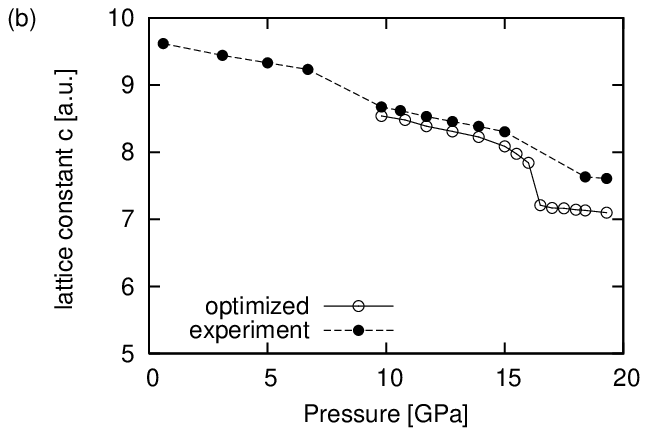}} & 
    \resizebox{60mm}{!}{\includegraphics{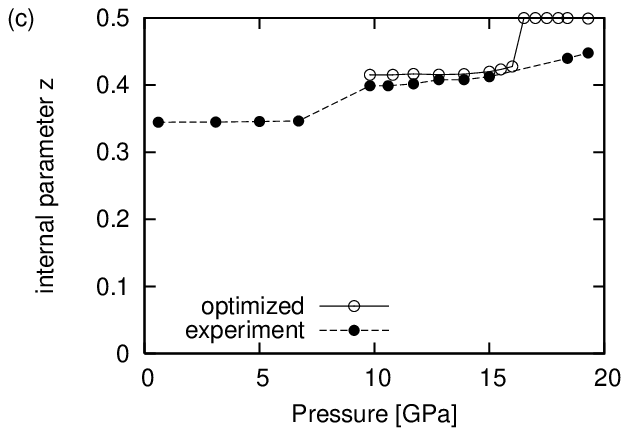}} 
   \end{tabular}
   \caption{Pressure dependence of lattice parameters: (a) lattice constant a,
   (b) lattice constant c, and (c) internal parameter z.
   The parameters are obtained by the structural optimization at each pressure,
   and they are indicated by open circles.
   The experimentally observed values\cite{str} are represented as closed circles.}
  \label{opt}

   \begin{tabular}{ccc}
    \includegraphics[width=8cm,height=6cm,clip]{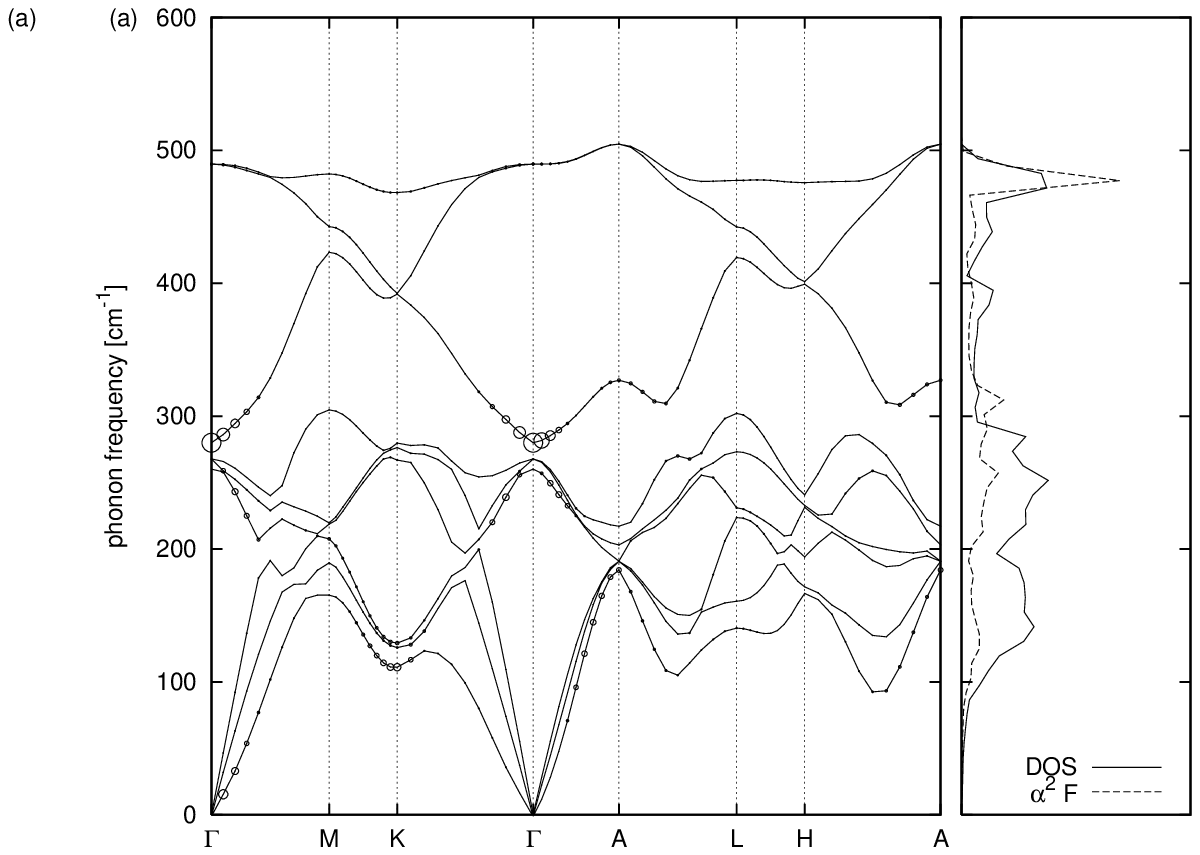} &
    \includegraphics[width=8cm,height=6cm,clip]{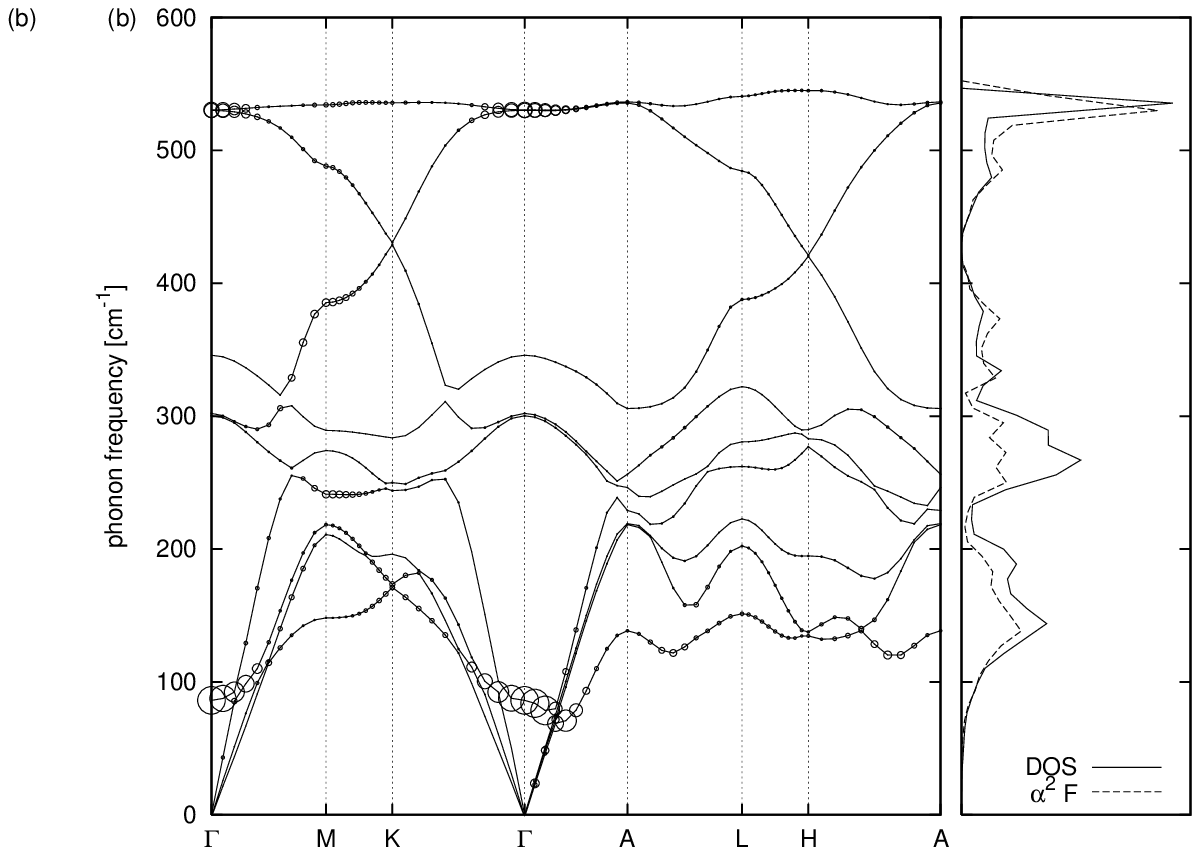} 
   \end{tabular}
   \caption{Phonon dispersion of CaSi$_2$.
   (a) The optimized trigonal structure at 10\,GPa.
   (b) The optimized AlB$_2$ structure  at 20\,GPa.
   Circles display partial electron-phonon interaction $\lambda_{\nu{\rm q}}$,
   which is explained in the text.}
\label{lambda}
 \end{center}
\end{figure}

\end{widetext}

\section{Results of the Calculations}

\subsection{Optimizations of the Structures in High Pressure Phases}

The lattice constants and the internal parameter were optimized by 
the constant-pressure variable-cell relaxation using
the Parrinello-Rahman method.\cite{PR}
In this calculation we used a $12\times12\times12$ ${\bf k}$-point grid in the Monkhorst-Pack grid
and set the energy cut-off for the wave functions at 16\,Ry and
that for the charge density is at 64\,Ry  for
the trigonal and the AlB$_2$ structure.
Though these values may be comparatively small, the accuracy is enough.
This is confirmed by the calculations with
larger energy cut-offs of 40\,Ry and 160\,Ry,
resulting in almost the same optimized structure.

In Fig.\ \ref{opt} we show the optimized lattice parameters.
In the pressure range from 10 to 17\,GPa,
the calculated lattice constants are in good agreement with the experimentally observed values.
The relative error of each constants, is less than 2\%.
Above 17\,GPa, however, the calculated lattice parameters disagree
with those of the AlB$_2$-like structure.
Especially, internal parameter $z$ becomes 0.5 in our calculation
while in the experiment\cite{str} it is less than 0.5 and does not reach that value.
Our result of optimization shows that CaSi$_2$ becomes
the AlB$_2$ structure above 17\,GPa.

\begin{figure}
\begin{center}
   \resizebox{60mm}{!}{\includegraphics[angle=-90]{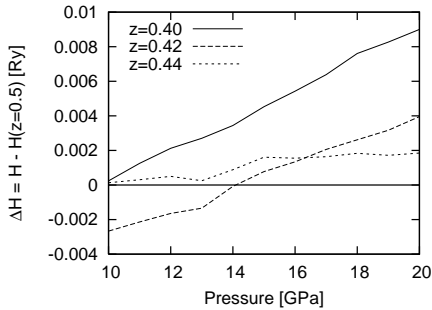}}
   \caption{The enthalpy curves of CaSi$_2$ in high pressure. 
   Each structure is determined with a fixed value of $z$.
   Enthalpy values relative to that of the AlB$_2$ structure(z=0.5) is given
   at each pressure.}
\label{fig:enth}
\end{center}
\end{figure}

This discrepancy is not due to the pseudopotential method adopted
in our simulation.
We checked the results
by the use of all electron methods:
The full potential linear muffin tin orbital 
method which is embodied in the packaged code developed
by S. Y. Savrasov and D. Y. Savrasov\cite{LMTO}
and the full potential linearized augmented-plane wave
method which is embodied in the WIEN2k code.\cite{LAPW}
We optimized the structure
with constant cell volume and obtained the same results.
In all methods, the calculated results indicate
that the AlB$_2$ structure is more stable than
trigonal structure at high pressure.
Here we note that the value of the pressure obtained
by the first-principles calculation could have an error in some cases.
In fact, the pressures of pure calcium estimated by the generalized
gradient approximation calculation,
which is the same method as the present one, are much lower than
those of experimental values.\cite{TI}
We expect that the AlB$_2$ structure will be observed at higher pressures
in the experiment.

To test stability of the AlB$_2$ structure,
we calculated the phonon frequency in the whole Brillouin zone.
The density functional perturbation theory was employed\cite{DFPT} 
for the phonon calculation, where $4\times4\times4$ ${\bf q}$-point grid 
was used.\cite{MP}
Phonon dispersion of each phase is shown in Fig.\ \ref{lambda}.
We observe that only real frequencies appear all over the Brillouin zone, 
which indicates that the AlB$_2$ structure is stable in the pressure range
higher than 17\,GPa.

In Fig.\ \ref{fig:enth},
we show pressure dependence of enthalpy of some atomic structures. 
Each structure is given by optimizing $c/a$ 
with fixing the value of $z$.
This graph indicates that
CaSi$_2$ abruptly transforms from the trigonal structure ($z=0.42$)
to the AlB$_2$ structure($z=0.5$).
In our simulation, thus no transition from the trigonal structure 
to the AlB$_2$-like structure (z=0.44) occurs.

\begin{widetext}
\begin{figure}[htpd]
\begin{center}
   \begin{tabular}{ccc}
   \resizebox{60mm}{!}{\includegraphics[angle=-90]{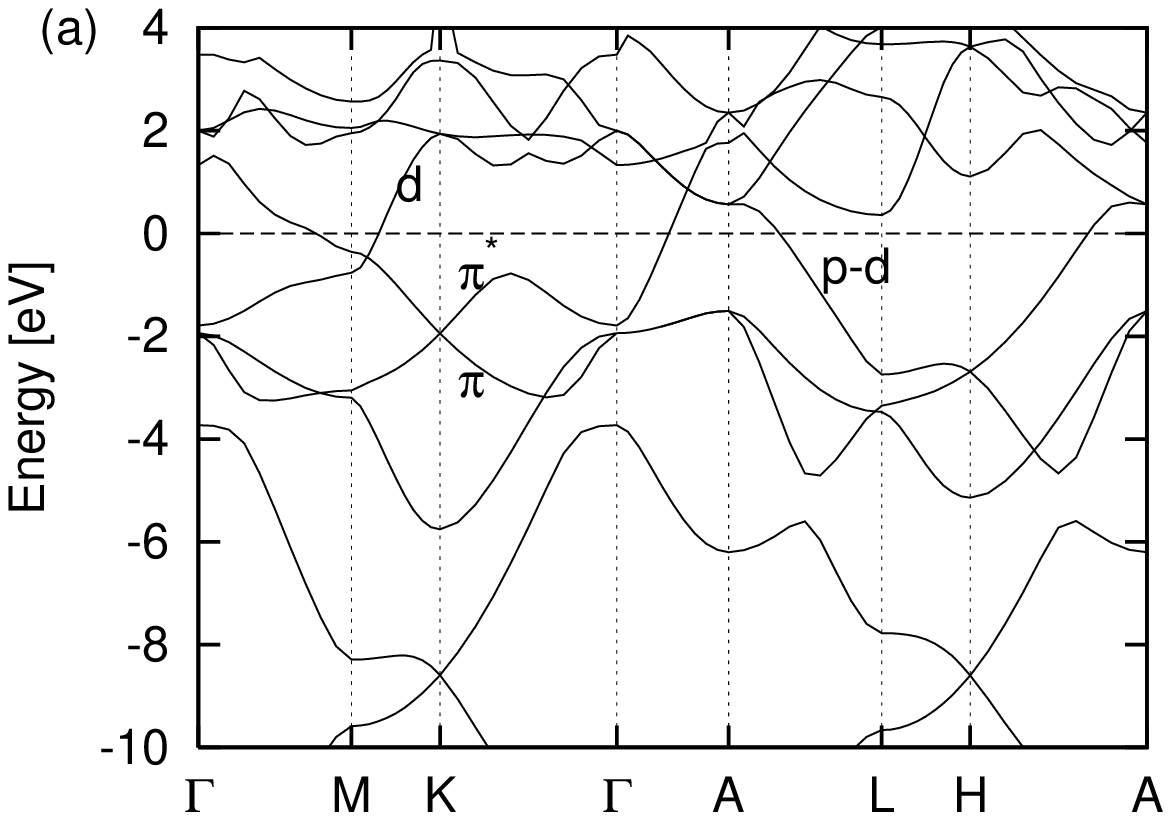}} &
   \resizebox{60mm}{!}{\includegraphics[angle=-90]{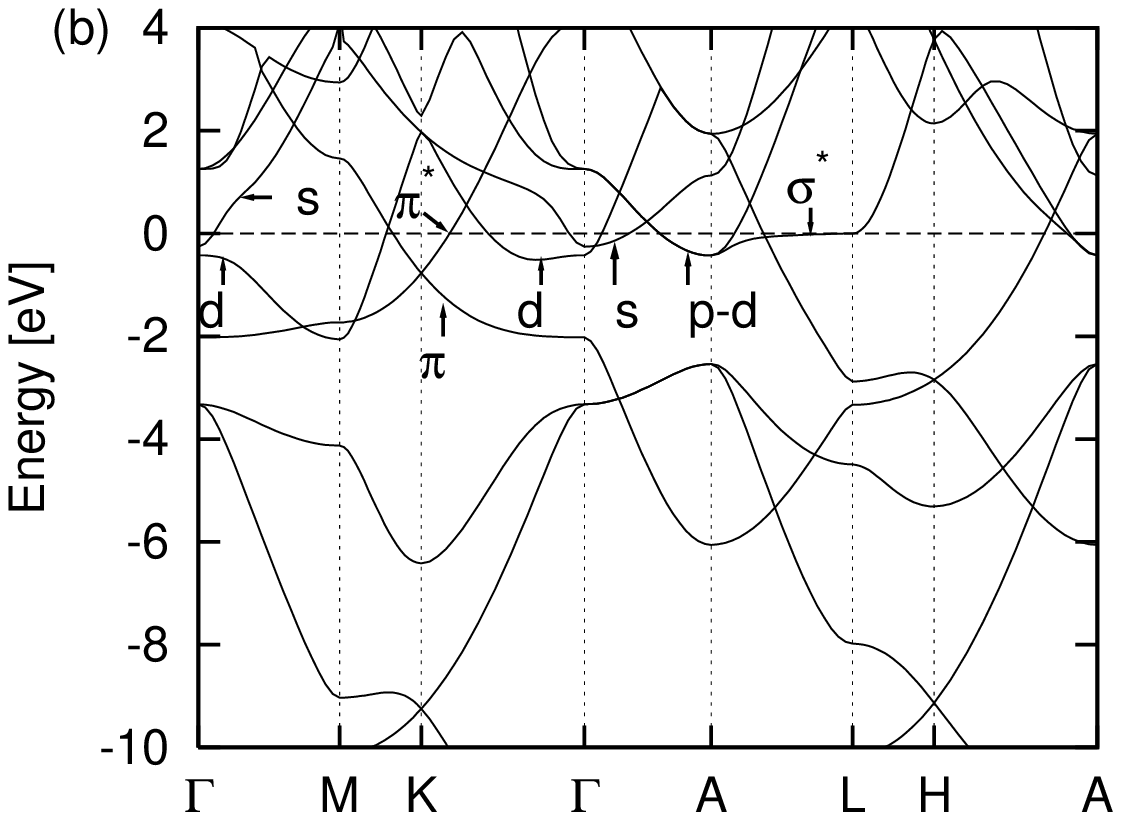}} &
   \resizebox{60mm}{!}{\includegraphics[angle=-90]{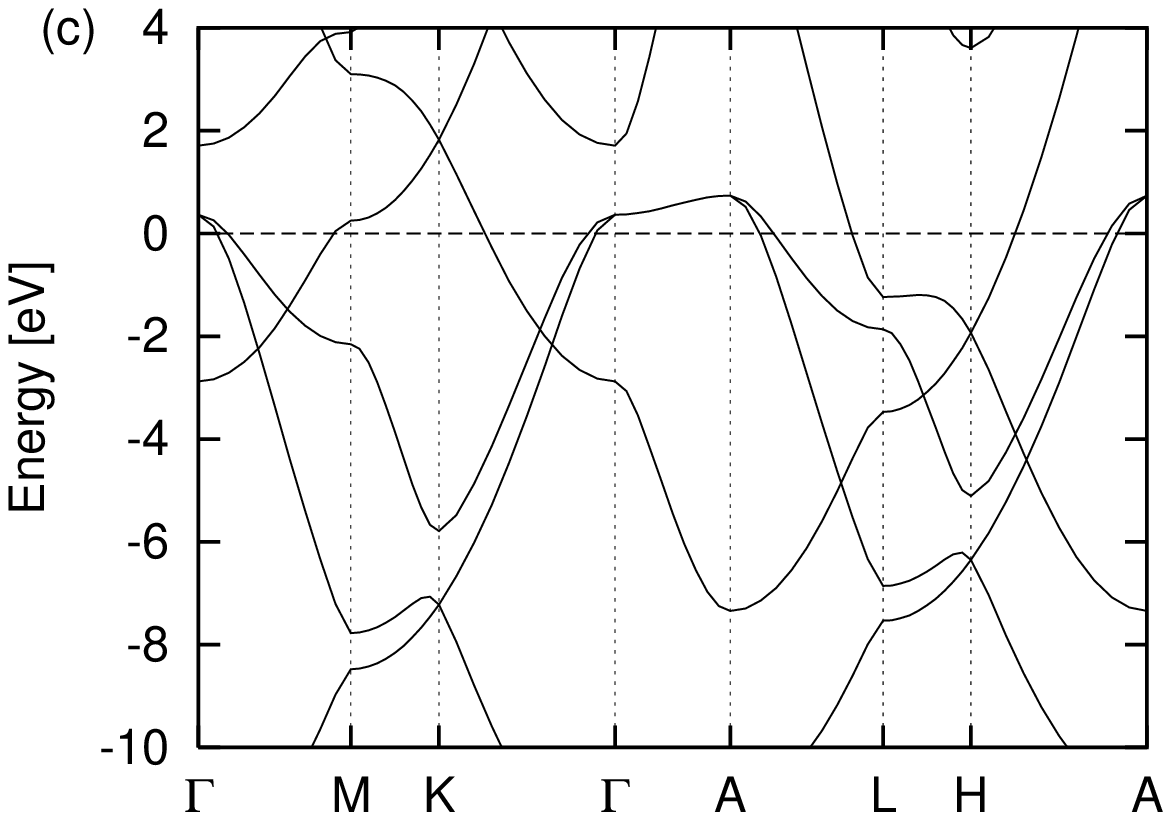}} \\\\
   \end{tabular}
   \caption{Electronic band dispersions of the optimized structures 
   of CaSi$_2$ in
      (a) the trigonal structure which is realized in the Phase III and
      (b) AlB$_2$ structure. (c) The electronic band structure of MgB$_2$.}
\label{band}
\end{center}
\end{figure}
\end{widetext}

\subsection{The Electronic Band Structure}

We now analyze the band structures of CaSi$_2$ obtained by the Kohn-Sham equations. 
(Fig.\ \ref{band}(a), (b))
The electronic band structure of CaSi$_2$ in the AlB$_2$ structures 
was studied in earlier works.\cite{KK,Satta}
Global features of dispersion relations are roughly the same, but
we can find some important differences between the present results
and the earlier ones.
This is partly due to optimization of the
crystal structure.
In the present work, we have fully optimized 
structure for the Phase III and the AlB$_2$ phase. 

In Fig.\ \ref{band} (a), the band structure of the trigonal structure
is shown.
The Si $p_z$ bands, Ca $d$-band and a $p$-$d$ hybridized band touch the Fermi level.
Since the Si planes are corrugated, $p_z$ bands of Si should rather be called "$\pi(\pi^*)$-like" band.

In Fig.\ \ref{band} (b),
we show the band structure of CaSi$_2$ in the AlB$_2$ structure. 
First, we can observe $d$-character in some hybridized bands
near the Fermi level.
One of those branches is seen at the $\Gamma$ point.
We should note that for a Ca compound in a high pressure phase
Ca $d$-orbitals often appear at the Fermi level.
Appearance of the $d$ character was pointed out theoretically for CaSi in 
the CuAu structure and CaSi$_3$ in the CuAu$_3$ structure.\cite{Weaver}
Second, the $\pi^*$ band lies also near the Fermi level.
At the $K$ point,
the crossing of the $\pi$ and $\pi^*$ bands is observed, 
which is a characteristic of the AlB$_2$ structure.
It looks that the electrons are doped in the $\sigma^*$ and $\pi^*$ 
bands of CaSi$_2$.
Third, the doubly degenerated $p$-$d$ hybridized bands at the $A$ point
are occupied.
Along the $A$-$L$ symmetry line, one of those bands becomes
almost dispersionless, which enhances the density of states around 
the Fermi level as shown below.

\begin{figure}[htpd]
\begin{center}
   \resizebox{60mm}{!}{\includegraphics[angle=-90]{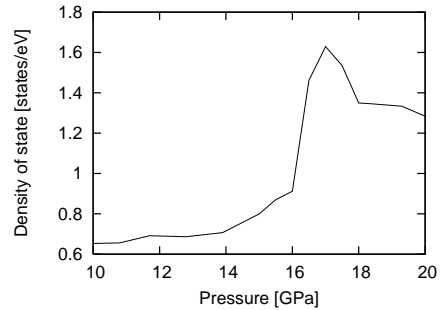}} 
   \caption{Pressure dependence of the density of states of CaSi$_2$ 
   at the Fermi level. Sudden increase happens when the structure changes to the AlB$_2$ structure at about 17GPa.}
\label{fig:dos} 
\end{center}
\end{figure}

We calculated the pressure dependence of
electronic density of states at the Fermi level(Fig.\ \ref{fig:dos}).
Through the structural transition,
density of states at the Fermi level 
increases  from 0.65 [state/eV] at 10\,GPa to 1.28 [state/eV] at 20\,GPa.
As shown in Fig.\ \ref{band}, in the AlB$_2$ structure,
the $s$, $d$ and $p$-$d$ bands go down to Fermi level
and make electron pockets.
This is the reason for the enhancement of the density of states.

Here we compare the electronic band structure of CaSi$_2$ with
that of MgB$_2$ (Fig.\ \ref{band} (c)).
Although these materials have a close resemblance with each other,
the band structure of MgB$_2$ has some important differences from that of 
the CaSi$_2$.
These AlB$_2$ structures have both $\pi$- and $\sigma$-bands 
of sp$^2$-hybridized orbitals. 
The $\sigma$ bands of MgB$_2$ looks to be partly hole-doped 
creating small two-dimensional hole pockets.\cite{band}
In CaSi$_2$, on the other hand, $\sigma$ bands are fully occupied
and a flat $\sigma^*$ band lies along the $A$-$L$ line around the Fermi level.
The MgB$_2$ is known to be a two-band system
which has $\pi$- and $\sigma$-bands.
CaSi$_2$, however, has additional bands around the Fermi level,
which are Ca $3d$ bands, Si $3s$ band, and $p$-$d$ hybridized bands.

\subsection{The Electron-Phonon Interaction and the Superconductivity}

\begin{figure}[htp]
 \begin{center}
   \begin{tabular}{c}
  \resizebox{50mm}{!}{\includegraphics{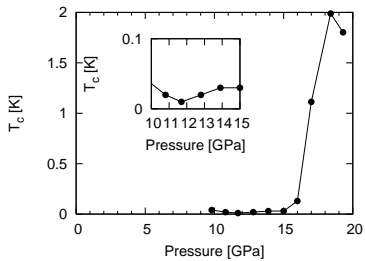}} \\\\
   \end{tabular}
   \caption{Estimation of pressure dependence of $T_{\rm c}$.
   For our simulation CaSi$_2$ takes the structure
   with the corrugated honeycomb network of Si atoms
   in the pressure range from 10 to 17\,GPa, and it transforms
   to the structure with the perfectly flat silicon network above 17\,GPa.
   Sudden enhancement of $T_{\rm c}$ is due to this structural transformation.}
   \label{grp.tc}
 \end{center}
\end{figure}

Assuming the phonon-mediated superconductivity,
we estimated the superconducting transition temperature
using the strong coupling theory,\cite{Eliashberg,AD} 
in which the electron-phonon matrix are calculated
by the density functional perturbation theory.
Our results of the $T_{\rm c}$ shows that
when the structural transition occurs,
$T_{\rm c}$ rises rapidly and reaches to a value
one order of magnitude larger than those in the low pressure phase.
The estimated superconducting transition temperature
are shown in Fig.\ \ref{grp.tc}.
In the trigonal structure,
calculated results are almost one-tenth
of the experimentally observed values,\cite{Tc-estimation} 
which is about 3$\sim$4\,K.
This discrepancy may be due to utilization of
an isotropic approximation in the Eliashberg theory.
In spite of the low values of the estimated $T_{\rm c}$,
we can discuss the pressure dependence of $T_{\rm c}$.

Let us examine the origin of the pressure dependence of $T_{\rm c}$
in our theoretical data.
According to the McMillan's formula,\cite{AD} 
$T_{\rm c}$ is given by three parameters; 
the electron-phonon coupling constant $\lambda$, 
the logarithmic average of the phonon frequency $\omega_{\log}$, 
and the Coulomb parameter $\mu^{\ast}$, in the following form. 
\[
T_{\rm c}=\frac{\omega_{\log}}{1.2}
\exp \left( -\frac{1.04(1+\lambda )}
{\lambda-\mu^{\ast}(1+0.62\lambda )} \right).
\]
Here $\lambda$ and $\omega_{\log}$ are obtained by 
the first-principle calculations using
the density functional perturbation theory.
As for $\mu^{\ast}$, we assume the value $\mu^{\ast}\sim$0.1
which holds for simple metals.
In the present substances, the critical temperature is mainly determined by
$\lambda$ and $\omega_{\log}$.
The table \ref{tabtc} shows $\lambda$ 
and $\omega_{\log}$ for the AlB$_2$ and the trigonal structures.
Of the two parameters,
the more significant increase is seen in $\lambda$.
While $\omega_{\log}$ drops about 10\%, the increase of the
$\lambda$ by a factor of 1.5 in the AlB$_2$ structure leads to
the enhancement of $T_{\rm c}$.

Here we analyze the electron-phonon interaction.
The parameter $\lambda$ is given explicitly as follows. 
\[\lambda \equiv 2\int_0^\infty d\omega \frac{\alpha^2 F(\omega)}{\omega},\]
using the electron-phonon spectral function, 
\[
\alpha^2F(\omega)=
\frac{
N(0)\sum_{{\bf k}\nu{\bf q}}|M_{{\bf k,k+q}}^{\nu{\bf q}}|^2 
\delta(\omega-\omega_{\nu{\bf q}})
\delta(\varepsilon_{\bf k})\delta(\varepsilon_{{\bf k+q}})}
{\sum_{{\bf kq}}\delta(\varepsilon_{\bf k})\delta(\varepsilon_{{\bf k+q}})}.
\]
Here $N(0)$ is the density of electronic states with a single spin component 
at the Fermi level, $\omega_{\nu{\bf q}}$ and $\varepsilon_{\bf k}$ are 
phonon and electron energies, and $M_{{\bf k,k+q}}^{\nu{\bf q}}$ is
the electron-phonon matrix elements. 
For the mode analysis, we introduce
partial electron-phonon interaction constant $\lambda_{\nu{\bf q}}$, defined by
\[
\lambda_{\nu{\bf q}}
=\frac{
2N(0)\sum_{{\bf k}}|M_{{\bf k,k+q}}^{\nu{\bf q}}|^2
\delta(\varepsilon_{\bf k})\delta(\varepsilon_{{\bf k+q}})}
{\omega_{\nu{\bf q}}\sum_{{\bf kq'}}\delta(\varepsilon_{\bf k})\delta(\varepsilon_{{\bf k+q'}})},
\]
from which the mean value is obtained as
$\lambda = \sum_{\nu {\bf q}} \lambda_{\nu{\bf q}}$.

Using $\lambda_{\nu{\bf q}}$, 
we find the most influential phonon mode 
for the superconductivity and $T_{\rm c}$, which give large contribution 
to the electron-phonon interaction parameter $\lambda$. 
The contribution is shown in Fig.\ \ref{lambda},
where $\lambda_{\nu {\bf q}}$ is shown by a circle
on each phonon dispersion, and the radius is proportional to
the contribution to electron-phonon interaction parameter $\lambda$.
This figure indicates that, in the AlB$_2$ structure,
the highest mode at the $\Gamma$ point is effective.
This mode is the E$_{2g}$ mode, in which the neighboring 
silicon atoms oscillate in the anti-phase within a Si plane. 
This feature is the same as that observed in the MgB$_2$,
in which the E$_{2g}$ mode is 
the key mode of the high-temperature superconductivity.\cite{Bohnen} 
We can see, however, appearance of the high-frequency peak in the phonon 
density of states and $\alpha^2F(\omega)$ due to the E$_{2g}$ mode 
of CaSi$_2$ is much similar to the result of AlB$_2$. 

In addition, we see another important mode in the AlB$_2$ structure, which is
the softened optical mode around the $\Gamma$ point.
This mode is the B$_{1g}$ mode corresponding to 
out-of-plane motion of silicon atoms, 
whose displacement makes the Si plane corrugated.  
Due to the softening, $\lambda_{\nu{\bf q}}$ has a large value.
However, the softening may reduce $\omega_{\log}$ given by
\[
\omega_{\log} 
=\exp \left(\frac{2}{\lambda}\int_0^\infty
d\omega\frac{\alpha^2F(\omega)}{\omega}\log\omega\right),
\]
and does not necessarily work to increase the transition temperature
as exemplified in iodine.\cite{MNKS}
In the case of CaSi$_2$ in the AlB$_2$ structure, we observe the high-frequency optical branch containing the $E_{2g}$ mode at the $\Gamma$ point. 
In this branch, phonon frequency becomes even higher 
than that of corresponding branch in the trigonal structure 
in the low pressure phase.
As a result, $\omega_{\log}$ is kept almost in the same order of magnitude
though in the low-frequency range the spectral function 
increases (Fig.\ \ref{lambda}). 
Consequently $T_{\rm c}$ is not decreased by the phonon softening.
This means that both B$_{1g}$ and E$_{2g}$ phonon modes contribute to
enhancement of electron-phonon interaction and $T_{\rm c}$.

\begin{table} [htpd]
\begin{center}
\caption{
Comparison of the electron-phonon interaction parameter $\lambda$
and the mean logarithmic frequency $\omega_{\log}$ between
a trigonal structure with the corrugated silicon plane at 10\,GPa
and AlB$_2$ structure with the perfectly flat silicon plane at 20\,GPa.
}
\begin{tabular}{ccc}
\hline \hline
structure  & $\lambda$  & $\omega_{\log}[K]$ \\ \hline 
 trigonal  & 0.27 & 300 \\ 
 AlB$_2$   & 0.41 & 280 \\
\hline \hline
\end{tabular}
\label{tabtc}
\end{center}
\end{table}

\section{Summary}

In this study, we found that
the AlB$_2$ structure
appears as the high-pressure phase of CaSi$_2$, 
and the enhancement of the superconducting transition temperature, 
$T_{\rm c}$, is expected in the AlB$_2$ phase.
The enhancement of $T_{\rm c}$ is due to
the enhancement in the electron-phonon interaction. 
If we assume that CaSi$_2$ has the phonon-mediated superconductivity,
the E$_{2g}$ and B$_{1g}$ phonon modes
play an important role in the enhancement of $T_{\rm c}$
through the transformation from the structure with corrugated Si plane
to the structure with the flat one. 

This work was supported by a Grand-in-Aid for scientific research 
(No.15GS0213, No.17064006, No.17064013, No.19051016), 
the 21st century COE program ``Core Research and Advanced Education Center for 
Materials Science and Nano Engineering'', 
and also by the next generation integrated nanoscience simulation software 
from the Ministry of Education, Culture, Sports, Science, and Technology, Japan.


\begin{thebibliography}{13}
\bibitem{Tc-estimation}
 For a test calculation, we estimated $T_{\rm c}$ of MgB$_2$ in the present scheme. The estimation tells $T_{\rm c} \simeq 25$K, which is close to the known value obtained with approximations same as the present study.
\end{thebibliography}
\end{document}